\documentclass[10pt]{iopart}
\usepackage{graphicx}
\usepackage{iopams}

\def \d{\mbox{d}}

\def \reals{I\!\!R}
\def \Minkowski{I\!\!M}
\def \dxx{\,\d^{2}\!x}
\def \dxxx{\,\d^{3}\!x}

\def \text{\mbox}
\def \Eq{equation }

\def \p{{\partial}}

\newcommand{\vol}[1]{{\text{vol}^{#1}}}

\def \bA{{\mathbf{A}}}
\def \bB{{\mathbf{B}}}
\def \bE{{\mathbf{E}}}
\def \bG{{\mathbf{G}}}
\def \bv{{\mathbf{v}}}

\def \bx{{\mathbf{x}}}
\def \bn{{\mathbf{n}}}
\def \be{{\mathbf{e}}}

\def \indi{i}
\def \indj{j}
\def \indk{k}

\begin{document}

\title[Third-order topological invariant]{Towards a third-order topological invariant for magnetic fields}

\author{Gunnar Hornig and Christoph Mayer}

\address{Theoretische Physik IV, Fakult\"at f\"ur Physik und Astronomie,\\
  Ruhr-Universit\"at Bochum, 44780 Bochum, Germany}

\eads{\mailto{gh@tp4.ruhr-uni-bochum.de}, \mailto{cm@tp4.ruhr-uni-bochum.de}}

%% Pacs: 95.30.Q, 02.40.M, 96.60.H
%%    Classical differential geometry  02.40.H,   Global
%%    differential geometry  02.40.M, solar magnetic fields
%%    96.60.H, Magnetohydrodynamics in astrophysics, 95.30.Q

\begin{abstract}

An expression for a third-order link integral 
of three magnetic fields is presented.
It is a topological 
invariant and therefore an invariant of
ideal magnetohydrodynamics.
The integral generalizes existing expressions for third-order
invariants which are obtained from the Massey triple product,
where the three fields are restricted 
to isolated flux tubes.
The derivation and interpretation of the invariant shows a close
relationship 
with the well-known magnetic helicity,
which is a second-order topological invariant.
Using gauge fields with an $SU(2)$ symmetry, helicity and the new
third-order invariant originate from the same identity, an identity
which relates the second Chern class and the Chern-Simons three-form.
We present an explicit example of three magnetic fields with non-disjunct
support.
These fields, derived from a vacuum Yang-Mills field with a
non-vanishing winding number, 
possess a third-order linkage detected by our invariant. 

\end{abstract}

\section{Introduction}
The topological structure of magnetic fields
is an important subject in plasma physics.
There, among other issues, it is related to 
the problem of stability of a plasma and to its energy content.
Fields with an enormous wealth of entangled, braided or knotted
field lines exist for example in the solar atmosphere. Note that the
topological complexity of these solar magnetic fields is only revealed,
if first, one takes into account that the observed loops anchored in
the photosphere  are closed by subphotospheric fields, and second,
that already simple toroidal equilibria contain many  different
knotted and linked field lines. The simplest examples are the so called
torus knots, which are formed by field lines where the quotient of the
number of windings around the core and the torus axis is rational.

In general, magnetic fields in plasmas are not static, but evolve
due to the motion of the plasma. The evolution of solar as well as
most astrophysical magnetic fields
is given in  good approximation by the induction equation of
ideal magnetohydrodynamics (IMHD) 
\begin{equation}
\frac{\partial }{\partial t} {\bf B} 
        - \nabla \times ({\bf v} \times {\bf B})  = 0,
        \label{idindeq}
\end{equation} 
which shows that the field can be considered as frozen-in with 
respect to the plasma velocity ${\bf v}$.
The approximation of IMHD which leads to this law is valid as 
long as the evolution does not lead to small scale structures,
 e.g.~thin current sheets. 

The ideal induction equation guarantees the conservation of the
topology of field lines under the flow of ${\bf v}$, i.e. every
linkage or knottedness of magnetic flux is preserved. Mathematically
speaking, the flow of ${\bf v}$ is a differentiable isotopy of the
space under consideration. It maps the field lines of ${\bf B}$
at time $t_0$ to a topologically equivalent set of field lines for
any later time $t> t_0$.
Let us note that, following the usual terminology in plasma physics,
the term  `topological equivalent' is used here in the
sense of a diffeomorphic isotopy.

\begin{figure}[b]
\begin{center}
\includegraphics[width=4.0cm]{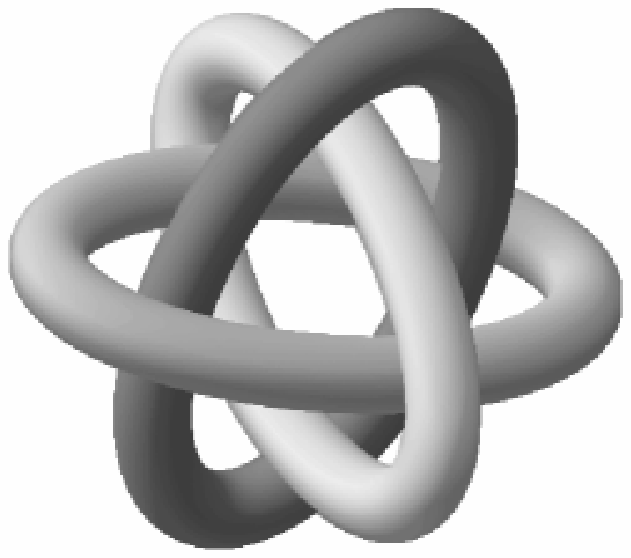}
\hspace{2cm}
\includegraphics[width=5.0cm,trim=0 6 0 0]{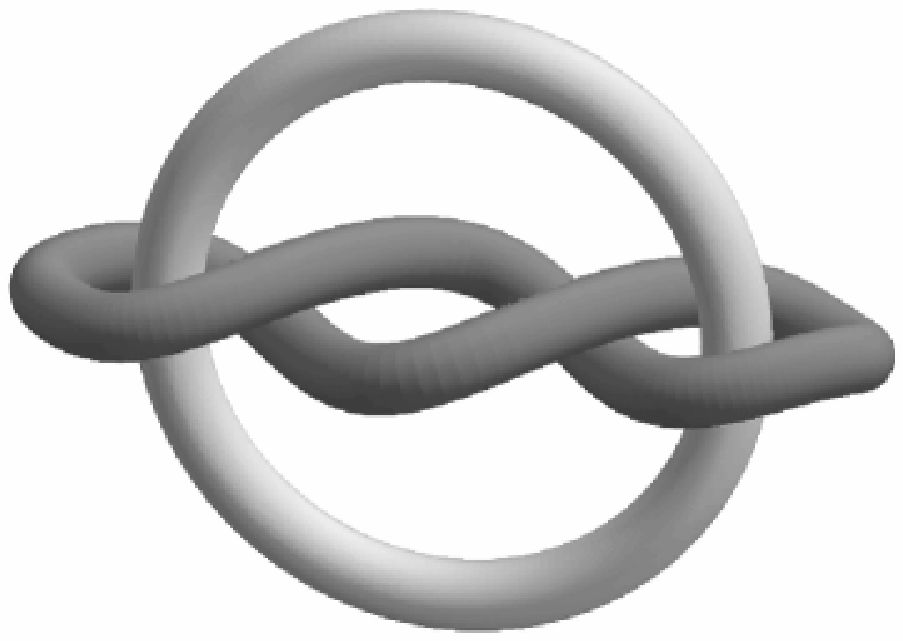}
\end{center}
\caption{The Borromean rings (left) and the Whitehead link (right).}
\label{borro}
\label{Whitehead}
\end{figure}

In order to describe the structure of magnetic fields, it is 
desirable to have measures of complexity.
These measures should be topological, i.e. they should be
invariant under an arbitrary isotopy of the magnetic field,
and therefore invariant under the ideal induction equation.
An example of a topological measure for 
magnetic fields is the  magnetic helicity,
a quantity which has attracted a great deal of 
attention in recent years (see e.g.~Brown \etal 1999).
Magnetic helicity, which measures the linkage of magnetic flux,
is only a lowest order topological measure.
It fails for instance to detect the interlocking of magnetic flux
tubes in form of the Borromean rings or the Whitehead link
(see Figure \ref{borro}).  
The total magnetic helicity of these configurations vanishes just 
as it does for three or two unlinked flux tubes. 
Both configurations 
must possess a higher order linkage or knottedness
of magnetic flux which is not detected by magnetic helicity.
This naturally raises the question whether corresponding higher order
measures similar to magnetic helicity exist which are sensitive to
these linkages. Here we would like to remark that the configurations  
shown in Figure \ref{borro} are highly idealized. In any real plasma we would not find this pure linkage but a mixture of  different types of linkages, each of which is to be  measured by a different integral.

In knot theory different invariants are known which distinguish
e.g. the Borromean rings or the Whitehead link from unlinked
rings.
The problem is that invariants used in physical applications, e.g. in
magnetohydrodynamics, should be expressed in terms of observable
quantities, in our setting in terms of the magnetic field $\bB$.
Up to now only the helicity,
which is related to the Gau\ss~ linking number,
has been formulated as an invariant for magnetic fields in a
satisfactory manner.
As was recognized first by Monastyrsky and Sasorov (1987)
and independently by Berger (1990) and Evans and Berger (1992),
the link invariants based on so-called higher Massey products
(see Massey 1958, 1969, Kraines 1966, Fenn 1983)
can be written as invariants applicable to magnetic flux tubes.
Similar to helicity, they only involve the magnetic fields and can be
expressed as volume integrals over the space in consideration.
Their disadvantage is that their usage is restricted
to magnetic fields confined to isolated flux tubes.
In addition,
these flux tubes must not possess a linkage lower than the linkage
which is measured.

In this paper we present a generalized third-order invariant for three
magnetic fields not confined to isolated flux tubes. 
In the case of isolated flux tubes this invariant coincides with the
invariant known from the Massey triple product.
Using gauge
fields in the context of an $SU(2)$ symmetry, the generalized invariant can be
shown to originate from the same equation as helicity.
Therefore, we will first recapitulate some basic facts about
 magnetic helicity before we turn to our main subject, the third-order
link invariant.

\section{Magnetic helicity}
Magnetic helicity of a field $\bB$ with arbitrary vector potential $\bA$ 
is defined as
\begin{equation}
  H(\bB) = \int_V \bA \cdot \bB \dxxx\,,  \quad\quad\quad 
  \bB \cdot \bn|_{\partial V} =0,                  \label{heldef}
\end{equation}
which can readily be
shown to be gauge invariant
if no magnetic flux crosses the boundary of the volume.
Since it is quadratic in magnetic flux it is often referred to
as a second-order topological invariant.
Magnetic helicity measures the total mutual linkage
of magnetic flux.
This interpretation can be motivated if we envisage a simple system of two
isolated and  closed flux tubes $U_1$ and $U_2$ with vanishingly 
small cross-section. The latter condition ensures that the
(self-)helicities of the flux tubes vanish.
Moffatt (1969) has shown that for this configuration the
helicity integral yields 
$H(\bB) = 2\,\phi_1\phi_2\,\text{lk}(U_1,U_2)$,
where $\text{lk}(U_1,U_2)$ is the Gau\ss~ linking number (Gau\ss~1867) 
of the two tubes and where $\phi_i$ is the magnetic flux in the tube
$U_i$.
Introducing an asymptotic linking number,
Arnol'd (1974) was able to extend this 
interpretation to the
generic case where field lines are not necessarily closed
(see also Arnol'd \& Khesin 1998, Part III, \S4).

Similar to {\em helicity} we can introduce the more general
{\em cross-helicity} of two magnetic fields $\bB_1$ and $\bB_2$. For a simply connected volume $V$ and provided that $\bB_1 \cdot \bn|_{\partial V} =\bB_2 \cdot \bn|_{\partial V} = 0$  we define 
\begin{equation}
   H(\bB_1,\bB_2) := \int_V \bA_1\cdot\bB_2 \dxxx
                  = \int_V \bB_1\cdot\bA_2 \dxxx . \label{crosshelicity}
\end{equation}
The boundary conditions ensure that this is again a gauge invariant quantity.
To see that the two integrals on the right-hand side
are equivalent, note that they differ only by a surface integral $\int
(\bA_1 \times \bA_2)\cdot {\bf n}  \dxx$.
This can be shown to vanish
using the equivalence of 
$\bB_i \cdot \bn|_{\partial V}=0$ with 
$\bA_i \times \bn|_{\partial V}=0$
in a 
certain gauge 
and for a simply connected volume $V$,
as proved in \ref{Equivboundaryconditions}.
Since both
integrals are gauge invariant this proves the equality for any gauge.

$ H(\bB_1,\bB_2)$ measures purely the cross linkage of flux among the
two fields.
Applied to our system of two isolated closed flux tubes with fields 
$\bB_1 + \bB_2 = \bB$ in the corresponding tubes this invariant yields
\begin{equation}
   H(\bB_1,\bB_2) = 2\,\phi_1\phi_2\,\text{lk}(U_1,U_2),
\end{equation}
which is now valid without the assumption of vanishingly small tube
cross-sections.

The significance of magnetic helicity arises from the fact that it is 
invariant in IMHD.
Using merely the homogeneous Maxwell equations we obtain
\begin{equation}
\p_t(\bA\cdot\bB)+\nabla\cdot\left(A_0\bB+\bE\times\bA\right) 
       \,=\, -2 \bE\cdot\bB,
\label{helicity-density} 
\end{equation}
which
describes the time evolution of helicity density $\bA\cdot\bB$.
Here $A_0$ is the electric potential and $\bE$ the electric field.
The term $A_0\bB+\bE\times\bA$ is to be interpreted as 
a {\em helicity current}, and
$-2\bE\cdot\bB$ as a {\em source term}.
In an ideal plasma, i.e. with $\bE + \bv\times\bB=0$, the source term
vanishes and the helicity current can be written as
\begin{equation}
   A_0\bB+\bE\times\bA = 
           A_0\bB+\bv\left(\bA\cdot\bB\right)-\bB\left(\bA\cdot\bv\right).
\end{equation}
Therefore \Eq (\ref{helicity-density}) takes the form:
\begin{equation}
   \p_t \left(\bA\cdot\bB\right)
       +\nabla\cdot\left(\bv\left(\bA\cdot\bB\right)\right) 
                        = \nabla\cdot\left(\chi\bB\right),
                        \label{helicity-conservation-equation}
\end{equation}
with $\chi=-A_0+\bA\cdot\bv$.
Elsasser (1956) already noticed that a 
particular gauge can be found such that $\chi=0$.
Using either this gauge or an arbitrary gauge together with the
boundary condition
$\bB\cdot\bn\big|_{\p V} = 0$
this last equation implies the conservation of helicity
in a comoving volume for an ideal plasma, since 
\begin{equation}
   \fl
   \frac{\d}{\d t}\int\limits_V \bA\cdot\bB \dxxx = 
   \int\limits_V \p_t \left(\bA\cdot\bB\right)
       +\nabla\cdot\left(\bv\left(\bA\cdot\bB\right)\right) \dxxx =
    \int\limits_{\p V} \chi\bB\cdot\bn \dxx \,= \,0.
    \label{helicity-conservation-proof}
\end{equation}
The invariance of integral (\ref{heldef}) was first stated by Woltjer (1958).

%%%%%%%%%%%%%%%%%%%%%%%%%%%%%%%%%%%%%%%%%%%%%%%%%%%%%%%%%%%%%%%%%%%%
%%%%%%
%%%%%%  3rd order invariant from Chern Simons three form
%%%%%%
%%%%%%%%%%%%%%%%%%%%%%%%%%%%%%%%%%%%%%%%%%%%%%%%%%%%%%%%%%%%%%%%%%%%
\section{A third-order invariant from the Chern-Simons three-form}
\label{section-third-order-invariant}
In this section we construct a third-order invariant
which, under conditions specified below, yields an invariant for three
magnetic fields. The derivation is based on some basic
knowledge in differential geometry found e.g.~in Frankel (1997). 

We have noted before that \Eq (\ref{helicity-density}) can be derived
purely from the homogeneous Maxwell equations.
Written in differential forms it reads   
\begin{equation}
 \d(A \wedge \d A)=  F\wedge F  \label{FwedgeF},
\end{equation}
where $A$ is the one-form potential of the field $F$.
The right-hand side of this equation is one of two 
(pseudo-) scalar Lorentz
invariants that can be constructed from the field tensor. We can
interpret this equation as a special case of a general result in the
theory of Chern forms, namely the exactness of the second Chern form
\begin{equation}
   \d\,tr(A\wedge \d A - i q \frac{2}{3}A \wedge A \wedge A) = tr(F\wedge F).
\label{Chern-Simons} 
\end{equation}
In this equation $A$ and $F$ are a matrix 
valued one- and two-form.
To be more precise they take their values
in the Lie-Algebra ${\frak g}$ of the structure group.
In Yang-Mills theory this is the symmetry group of the 
interaction under consideration, with coupling constant $q$.
On the vector space ${\frak g}$ the trace defines a
natural scalar product.
For $\omega$, $\theta \in {\frak g}$,
$\tr(\omega\wedge\theta)= \omega_{ij}\wedge \theta_{ji}$, where the
indices denote matrix components.

Equation (\ref{Chern-Simons})
holds for an arbitrary, not necessarily Abelian, field
strength $F = \d A - i q A \wedge A$. The three-form on the left-hand side
is known as the Chern-Simons three-form. For the case of electrodynamics,
i.e. for the Abelian structure group $U(1)$, $F$ is given by
$F=\d\, A$, since $A \wedge A $ vanishes and \Eq (\ref{Chern-Simons})
reduces to  (\ref{FwedgeF}).  In the non-Abelian case
\Eq (\ref{Chern-Simons})
splits into a real and imaginary part
\begin{eqnarray}
   \tr(\d A \wedge \d A) = \d\tr(A\wedge \d A), \label{identity-cs-real}\\
   2\tr(A \wedge A \wedge \d A ) 
   = \d\tr(\frac{2}{3}A \wedge A \wedge A) .\label{identity-cs-imaginary}
\end{eqnarray}
As we will see in the following  a third-order invariant can be
derived from identity (\ref{identity-cs-imaginary}) for the special case
of the structure group $SU(2)$.

Working with  an $SU(2)$ structure group it is appropriate to choose the
Pauli matrices $\sigma_i$, $i$=$1,2,3$, as a 
basis for the Lie Algebra $su(2)$.
All results, however, are independent of this
choice. The gauge potential $A$ and field strength $F$ now have 
three components
\begin{eqnarray}
   A = \sigma_\indi A^\indi, \qquad\qquad   F = \sigma_\indi F^\indi, 
                             \label{SU-two-AandF} 
\end{eqnarray}
where the summation convention over repeated indices is assumed.
Let us note that in the following 
we will refer to these fields as Yang-Mills fields, 
although they do not necessarily satisfy the Yang-Mills equation.
Using the identities for Pauli matrices
\begin{eqnarray}
    \sigma_j\sigma_k = i\;\epsilon_{jkl}\;\sigma_l + \delta_{jk}
    \mathbf{1}
    \qquad \text{and} \qquad
    \sigma_j\,\sigma_k\,\sigma_l = i\,\epsilon_{jkl} \mathbf{1},  
  \label{Pauli-matrices-identities}
\end{eqnarray}
equations (\ref{identity-cs-real}) and (\ref{identity-cs-imaginary}) read
\begin{eqnarray}
  \d A^\indi \wedge \d A^\indi
               = \d (A^\indi \wedge \d A^\indi),\\ 
 \epsilon_{\indi \indj\indk }\, \d A^\indi \wedge A^\indj \wedge A^\indk = 
 \frac{1}{3}\, \d (\epsilon_{\indi \indj\indk }
                      A^\indi \wedge A^\indj \wedge A^\indk ).
\end{eqnarray}
If we now interpret the three components of the Yang-Mills potential
$A$ as
three independent potentials of three electromagnetic fields 
$F_{EM}^\indi = \d A^\indi$,
the first identity states the helicity conservation (in IMHD)
for the sum of the self-helicities of the three individual
fields $F_{EM}^\indi$, similar to the electrodynamic case.
The second identity is new.
For convenience we introduce the two-form $G^1 = A^2 \wedge A^3$,
here on $\Minkowski^4$,
and cyclic permutations of it. 
Then we can write the second identity as
\begin{equation}
   \frac{1}{3}\d(A^i\wedge G^i) = F_{EM}^i \wedge G^i.\label{Second-identity}
\end{equation}
To complete the analogy of this equation with
\Eq (\ref{helicity-density}) we have to rewrite it in the language of
three-vectors. Therefore we represent the one-form $A^i$ by the time
component $A_0^i$ and the three-vector $\bA^i$ of the corresponding
four-vector. The two-form $G^1$ is identified with the vector pair
$(\bG_E^1,\bG_B^1) = (A_0^3\bA^2-A_0^2\bA^3, \bA^2\times\bA^3)$,
equivalent to the identification of  $F_{EM}$ with 
the three-vector pair ($\bE$,-$\bB$).
Cyclic permutations immediately lead to corresponding 
pairs for $G^2$ and $G^3$.
Using these conventions, the left- and right-hand side of
\Eq (\ref{Second-identity}) read respectively 
\begin{eqnarray}
   \frac{1}{3}\d(A^i\wedge G^i)
         &= -\frac{1}{3}\left(\p_t(\bA^i\cdot\bG_B^i)
                     +\nabla\cdot(A_0^i\bG_B^i+\bA^i\times\bG_E^i)\right)
                                        \vol{4}\nonumber\\
      &= -\left(\p_t(\bA^1\cdot\bG_B^1)
                      +  \nabla\cdot(A_0^i\bG_B^i)\right)\vol{4}
                    \nonumber
\end{eqnarray}
and
\[  F_{EM}^i \wedge G^i = (\bE^i\cdot \bG_B^i-\bB^i\cdot\bG_E^i) \vol{4}    
               = (\bE^i\cdot \bG_B^i-A_0^i\nabla\cdot\bG_B^i)\vol{4}.
\]
Thus, identity (\ref{Second-identity}) is equivalent to
\begin{equation}
   \p_t(\bA^1\cdot\bG_B^1)+\nabla\cdot(A_0^i\bG_B^i)
    = -\bE^i\cdot \bG_B^i+A_0^i\nabla\cdot\bG_B^i,
    \label{third-order-density-evolution}
\end{equation}
which shows a similar structure as \Eq (\ref{helicity-density})
in the case of helicity.
It describes the time evolution of the density 
$\bA^1\cdot\bA^2\times\bA^3$
with its current $A_0^i\bG_B^i$
and source term $-\bE^i\cdot \bG_B^i+A_0^i\nabla\cdot\bG_B^i$. This is
the basis for the following theorem.\\

\noindent
{\bf Theorem: }{\em 
Let $\bB^1$, $\bB^2$ and $\bB^3$ 
be three magnetic fields with potentials
$\bA^i$ satisfying $\bE^i+\bv \times\bB^i=0$. The integral over 
a volume $V\subset\reals^3$}
\begin{equation}
   H^{(3)}(\bB^1,\bB^2,\bB^3) := \int_{V} \bA^1\cdot\bA^2\times\bA^3\dxxx
   \label{H-3-continuous-integral}
\end{equation}
{\em
is a gauge invariant, conserved quantity, if
\begin{enumerate}
\item the potentials obey $\nabla\cdot(\bA^i\times\bA^j)=0$ for all
  $i,j=1,2,3$, \label{H-3-condition-1}
\item the potentials obey the boundary condition 
  $\bA^i\times\bn|_{\p V}=0$ for $\bn$ 
  being the normal vector to the boundary of the integration volume $V$.
  \label{H-3-condition-2}
\end{enumerate}
}

\noindent

\noindent
{\em Proof}:
Let us first remark that
condition (\ref{H-3-condition-2}) of the theorem 
implies $\bB^i\cdot\bn|_{\p V}=0$ as shown in \ref{Equivboundaryconditions}.
It is  therefore consistent with condition 
(\ref{H-3-condition-1}) since 
$0 = (\bB^i\cdot\bA^j-\bA^i\cdot\bB^j)|_{\p V} 
   = \nabla\cdot(\bA^i\times\bA^j)|_{\p V} $.  
Moreover, we show in \ref{Equivboundaryconditions} that for a simply
connected volume with $\bB^i\cdot\bn|_{\p V}=0$ condition
(\ref{H-3-condition-2}) can always be satisfied.
In order to prove the invariance of $H^{(3)}$ we observe that in
an ideal dynamics \Eq (\ref{third-order-density-evolution}) can be written as
\begin{eqnarray}
  \fl\nonumber
   \p_t(\bA^1\cdot\bA^2\times\bA^3)
     \,+\, &\nabla\cdot\left( \bv(\bA^1\cdot\bA^2\times\bA^3) \right) =\\ 
       &\nabla\cdot\left( \frac{1}{3}\bv(\bA^i\cdot\bG_B^i)-A_0^i\bG_B^i\right)
       + (A_0^i-\bA^i\cdot\bv)\nabla\cdot\bG_B^i ,\nonumber
\end{eqnarray}
where the last term vanishes due to condition (\ref{H-3-condition-1}).

Integrating 
over the volume $V$ yields the total time
derivative of $H^{(3)}$ on the left-hand side,
while the right-hand side can be converted into a surface integral,
analogous to \Eq (\ref{helicity-conservation-proof}) in the case of helicity.
The surface integral vanishes since condition 
(\ref{H-3-condition-2}) of the theorem
implies $\bG_B^i=0$ on the boundary of the volume $V$.
This, together with the gauge invariance of $H^{(3)}$
shown in \ref{appendix-H-3-gauge-invariance},
completes the proof.

\section{Interpretation of the invariant}
It is interesting to note that the new 
third-order invariant comes on an equal footing as the
conservation of helicity, since both invariants where derived from
the same identity (\ref{Chern-Simons}).
However, contrary to the conservation of helicity, the third-order
integral cannot be applied to a single magnetic field,
but requires  a
triplet of fields. Thus we have to interpret this integral in the
sense of the cross-helicity rather than the total helicity.
A forthcoming paper will deal with the question of
how 
a single magnetic field 
might be split into
a triplet of
fields with the required properties $\nabla\cdot\bG^i=0$,  thereby
linking the given {\it cross} third-order invariant and  a {\it total} 
third-order invariant.

There is 
another way of looking at the new third-order invariant.
By writing
\begin{equation}
   H^{(3)}(\bB^1,\bB^2,\bB^3) = \int_{V} \bA^1\cdot\bG_B^1
   \dxxx = H(\bB^1, \bG_B^1) \ ,
\end{equation}
the integral 
is to be interpreted as 
the cross-helicity of the two divergence-free fields
$\bB^1$ and $\bG_B^1$.
Note that the boundary conditions for the cross-helicity, namely
$\bB^1\cdot\bn=0$ and $\bG_B^1\cdot\bn=0$, are fulfilled due to
condition (\ref{H-3-condition-2}) of the theorem.
From this new interpretation the condition
$\nabla\cdot\bG_B^1=0$ is an obvious requirement 
analogous to $\nabla\cdot\bB=0$.
Furthermore, the symmetry of $H^{(3)}$ leads to 
\begin{equation}
   H(\bB^1, \bG_B^1) = H(\bB^2, \bG_B^2) = H(\bB^3, \bG_B^3),
   \label{H-3-cross-helicity}
\end{equation}
which reveals the additional conditions
$\nabla\cdot\bG_B^2=0$ and $\nabla\cdot\bG_B^3=0$.
Let us note that 
this interpretation does not simplify the direct 
calculation of the third-order invariant.
It is still necessary
to determine the fields $\bG_B^i$
which are not independent of
the chosen representatives $\bA^i$.

Third-order linking integrals for magnetic fields 
have been constructed from the Massey
triple product already by Monastyrsky and Sasorov (1987), Berger (1990)
and also Ruzmaikin and Akhmetiev (1994). However,
these constructions are limited to cases where the three fields
are confined to three isolated and mutually unlinked flux tubes with
disjunct support.
It is in fact easy to
see that for this special case their invariants coincide with the integral
(\ref{H-3-continuous-integral}) given above.
An explicit proof is given in \ref{Equivalentintegrals}.
In particular, it is worth noting that for fields with
mutually disjunct support the condition $\nabla\cdot \bG^i_B=0$ implies that
the cross-helicity of all pairs of fields vanishes, i.e.~their flux tubes
have to be mutually unlinked.

For a set of three arbitrary magnetic fields $\nabla\cdot \bG^i_B=0$ cannot
always be satisfied.
To show that there are examples
for which this can be satisfied,
beyond the cases of three fields with mutually disjunct
support,
we give an explicit example in the next section.

%%%%%%%%%%%%%%%%%%%%%%%%%%%%%%%%%%%%%%%%%%%%%%%%%%%%%%%%%%%%%%%%%%%
%%%%%  Example of three continuous magnetic fields ...
%%%%%%%%%%%%%%%%%%%%%%%%%%%%%%%%%%%%%%%%%%%%%%%%%%%%%%%%%%%%%%%%%%%

\section{Example of three magnetic fields with a 
third-order linkage}

In this section we want to give an example of three
magnetic fields not confined to flux tubes,
which firstly allow for one-form potentials that obey 
$\d(A^i \wedge A^j)=0$ and where secondly the integral invariant
(\ref{H-3-continuous-integral}) yields a non-trivial result. 
The existence of such an example proves that 
the new invariant is indeed a
generalization of the third-order invariant derived from a
Massey triple product which was applicable 
merely to unlinked flux tubes.
The fields we construct show an extraordinary high symmetry. For
this reason they are interesting in their own right.

The idea to construct three fields $A^i$ on $\reals^4$
which obey $\d(A^i \wedge A^j)=0$ comes from Yang-Mills theory:
An $SU(2)$ Yang-Mills field
\begin{equation}
   F = \d A - iq A \wedge A
\end{equation}
can, in view of \Eq (\ref{SU-two-AandF}) and
the identities for Pauli matrices (\ref{Pauli-matrices-identities}),
be written as
\begin{equation}
   F^\indi = \d A^\indi + q\,\epsilon_{\indi\indj\indk}A^\indj\wedge
   A^\indk. \label{F-alpha-vanishes}
\end{equation}
By taking the exterior derivative of $F$,
\begin{equation}
   \d F^\indi = q\,\epsilon_{\indi\indj\indk}\d(A^\indj\wedge A^\indk),
   \label{dF-su2}
\end{equation}
we immediately observe that $\d F=0$ is a sufficient condition for 
all $\d(A^\indi \wedge A^\indj)$ to vanish.
In the special case of a vacuum Yang-Mills field, i.e. $F=0$, the
requirement $\d F=0$ is trivially fulfilled.
If we now reinterpret the three components of the Yang-Mills potential
as potentials of three independent
magnetic fields, we have constructed an example field configuration to
which the invariant (\ref{H-3-continuous-integral}) can be 
applied.

\subsection{Yang-Mills potentials of a vacuum field with non-vanishing
  winding number}

An $SU(2)$ Yang-Mills vacuum field 
is now constructed
on a time slice $\reals^3$ of $\reals^4$
using the mapping $g: \reals^3 \rightarrow SU(2)$ (see e.g.~Frankel 1997, Itzykson and Zuber 1980)
\begin{equation}
   g(\bx) \,=\, \text{exp}\left[\frac{i\pi \; x^j \sigma_j}
                          {\sqrt{||\bx||^2 + \lambda^2}}\right].
                        \label{g-original}
\end{equation}
Interpreted as a gauge transformation of an $SU(2)$ classical vacuum,
i.e. with vanishing
connection  $\omega = 0$,
$g(\bx)$ gives rise to the pure gauge connection
\begin{equation}
   \omega = g^{-1}(\bx)\d g(\bx),\label{pure-gauge-omega}
\end{equation}
and the Yang-Mills potential one-form $A = \frac{i}{q}\omega$ reads
\begin{equation}
 A^j \sigma_j = \frac{i}{q}g^{-1}(\bx)\d g(\bx).
         \label{pure-gauge-omega-A} 
\end{equation}
At this point we want to remark that
the vacuum winding number of $\omega$,
which is defined to be the degree of the map $g$,
is $W(g)=1$.
An important consequence of this non-trivial winding number will be a 
non-trivial value of the invariant $H^{(3)}$, as can be seen in
equations (\ref{H-3-in-example}) and (\ref{H-3-and-W}) below.

In order to explicitly calculate the one-form potential $A$ 
given by the last expression we use
that $g(\bx)$, as an element of $SU(2)$, has the form
\begin{eqnarray}
   g(\bx) = \exp\left(\frac{1}{2i}\sigma_j n^j \right)
          = \cos\left({\frac{f}{2}}\right) 
                  -i \sigma_j \hat{n}^j \sin\left({\frac{f}{2}}\right),
                  \label{g-in-exp-form}
\end{eqnarray}
where $\bn = f {\hat\bn}$ 
and where $\hat{\bn}$ is a unit vector in $\reals^3$ with 
coordinate components $\hat{n}^j = x^j/||\bx||$. 
A comparison with \Eq  (\ref{g-original}) shows
\begin{eqnarray}
   f &= - \frac{2 \pi \,||\bx||}{\sqrt{||\bx||^2 + \lambda^2}}. \label{definition-f}
\end{eqnarray}
Substituting \Eq  (\ref{g-in-exp-form}) into (\ref{pure-gauge-omega-A})
we obtain after some calculation
\begin{equation}
   A^\indi =  \frac{1}{2q}\left[
               \hat{n}^\indi\,\d f + \d \hat{n}^\indi \,\sin{f} 
               +
               \epsilon_{\indi\indj\indk}\,\hat{n}^\indj\,\d\hat{n}^\indk 
               (\cos{f}-1)\right]. \label{Yang-Mills-potential-components}
\end{equation}

\subsection{Magnetic fields constructed from Yang-Mills potentials}

As mentioned above we now interpret the three components of the 
non-Abelian pure gauge Yang-Mills potential $A$ as potentials of 
three independent, non-vanishing magnetic fields.
It is sufficient to consider only one of the three potentials $A^i$,
since due to $\hat{n}^i = x^i/||\bx||$ and the cyclic symmetry 
of \Eq  (\ref{Yang-Mills-potential-components})
in the indices $i,j,k$ 
all three fields can be obtained from just one field by rotations
that map the $x^i$-axes on one another.
From \Eq  (\ref{Yang-Mills-potential-components}) we calculate 
the vector potential $\bA^3$ in spherical coordinates
$   x = r \sin\vartheta \cos\varphi$,
$   y = r \sin\vartheta \sin\varphi$, and 
$   z = r \cos\vartheta$.
Using unit basis vectors $\be_r,\be_\vartheta, \be_\varphi $ and fixing a value for
the ``coupling constant'' of $q=2$, we find
\begin{eqnarray}
\fl
   \bA^3 \,=\, 
   \frac{-2\pi \cos\vartheta }{(1+r^2)^{3/2}}\,\be_r \,+\,
   \frac{\sin\vartheta}{r}\sin\left(\frac{2\pi r}{\sqrt{1+r^2}}\right)\,
   \be_\vartheta
   \,-\,\frac{2 \sin\vartheta}{r}\sin^2\left( \frac{\pi r}{\sqrt{1+r^2}}\right)
   \be_\varphi. \label{Equation-Ai}
\end{eqnarray}
The corresponding magnetic field can be calculated from $\bB^3 =
 \nabla\times\bA^3$. It is easy to check that the fields $\bB^i$ are well
 defined and scale as
$ || \bB^i(\bx) || \rightarrow  4\pi^2$ for $r\rightarrow 0$ and as
$ || \bB^i(\bx) || \rightarrow  r^{-4}$ for $r\rightarrow \infty$.
Hence, they have no singularity and decay sufficiently fast for large radii.

Let us note that the fields $\bB^i$ are highly symmetric and
similar. Looking at the vector potential $\bA^3$ we observe that
it is independent of the variable $\varphi$, therefore $\bB^3$ is invariant
under rotations leaving the $x^3$-axis fixed.
Since the potentials (\ref{Yang-Mills-potential-components}) are
cyclic in $i,j,k$ it follows that each $\bB^i$ field 
is invariant under rotations about the $x^i$ axis.
We have pointed out before that a rotation that maps the 
Euclidian basis vector field $\be_{1}$ to $\be_{2}$ also maps $\bB^1$ to
$\bB^2$ etc. 
Furthermore,  $\bB^3$ is similar to the total field
$\bB = \bB^1+\bB^2+\bB^3$ in the following sense:
Let $R$ be a rotation that maps the Euclidian basis vector $\be_{3}$
to the vector $\sqrt{3}\,R(\be_{3}) = \be_{1}+\be_{2}+\be_{3}$, then
$\bB = \sqrt{3}\,R(\bB^3)$.

The magnetic fields $\bB^i$ are only of interest for us if their 
third-order invariant 
does not vanish.
Explicitly calculating $\bA^i$ for $i=1,2,3$ we find,
using the main Theorem,
\begin{eqnarray}
\fl
H^{(3)}(\bB^1,\bB^2,\bB^3) 
   &= \int_0^\infty \frac{4\pi}{r^2(1+r^2)^{3/2}}
      \left(\cos(\frac{2\pi r}{\sqrt{r^2+1}})-1\right)\,dr
   = -16 \pi^2 \label{H-3-in-example}.
\end{eqnarray}
The fact that this integral is non-vanishing proves, that the
constructed invariant cannot only be applied to all cases for which
we where able to calculate the already existing invariant, i.e.~to
three mutually unlinked flux tubes, but also to examples of triples of 
fields not having
disjunct support.
It is thus a true
generalization of the existing invariant known from the Massey
triple product.

As we have pointed out before, $H^{(3)}$ is
related to the vacuum winding number $W(g)$ of the connection 
$\omega=g^{-1}\d g$. We easily find (see also Frankel 1997) 
\begin{eqnarray}
 W(g)  &:= \nonumber
 \frac{1}{24\pi^2}\int_V tr(g^{-1} d g \wedge g^{-1} d g \wedge g^{-1} d g)\\
         &= 
   - \frac{1}{96\pi^2}\int_V\epsilon_{\indi\indj\indk}
                  A^\indi \wedge A^\indj \wedge A^\indk
  \label{H-3-and-W} \\
         &= -\frac{1}{16\pi^2} H^{(3)}(\bB^1,\bB^2,\bB^3), \nonumber 
\end{eqnarray}
where the trace term is usually 
referred to as the Cartan three-form on $SU(2)$.

In the general case the cross-helicities of three magnetic
fields, for which we are able to find potentials such that 
$\nabla\cdot(\bA^i\times\bA^j)=\bB^i\cdot\bA^j-\bA^i\cdot\bB^j= 0$,
do not have to vanish.
In our example we can easily verify that they do vanish, i.e.
$H(\bB^i,\bB^j) = 0$ for $i\ne j$.
Of more interest are the three non-trivial self-helicities.
If a triple of magnetic fields is derived from 
a Yang-Mills vacuum, \Eq  (\ref{dF-su2}) together with $F=0$ implies
\[
  \d A^\indi+\frac{1}{2}\,\epsilon_{\indi\indj\indk}A^\indj\wedge A^\indk = 0.
\]
Using the definitions 
$ B^\indi = \d A^\indi$
and 
$G^\indi = A^\indj \wedge A^\indk$
we find for $(\indi,\indj,\indk)$ cyclic
\[
   B^\indi = -A^ \indj \wedge A^\indk = - G^\indi.
\]
Thus, for $i=1,2,3$, we observe that
\[
   H(\bB^i,\bB^i) =  \int_V \bA^i\cdot\bB^i \dxxx 
                  = -\int_V \bA^i\cdot\bG^i \dxxx = -H^{(3)}(\bB^1,\bB^2,\bB^3).
\]
Therefore the self-helicities are equal to the value of the third-order
invariant. This is a peculiarity of all magnetic field triples 
derived from an $SU(2)$ Yang-Mills vacuum.

\begin{figure}[b]
\begin{minipage}[b]{0.45\textwidth}
\begin{center}
\includegraphics[width=\textwidth]{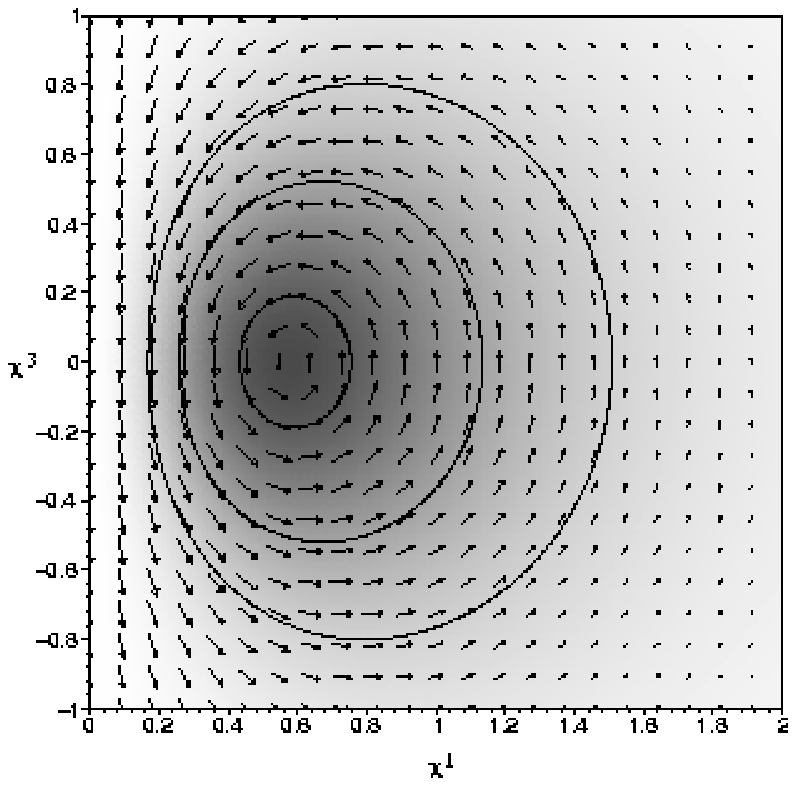}
\end{center}
\end{minipage}
\begin{minipage}[b]{0.45\textwidth}
\begin{center}
\includegraphics[width=0.82\textwidth]{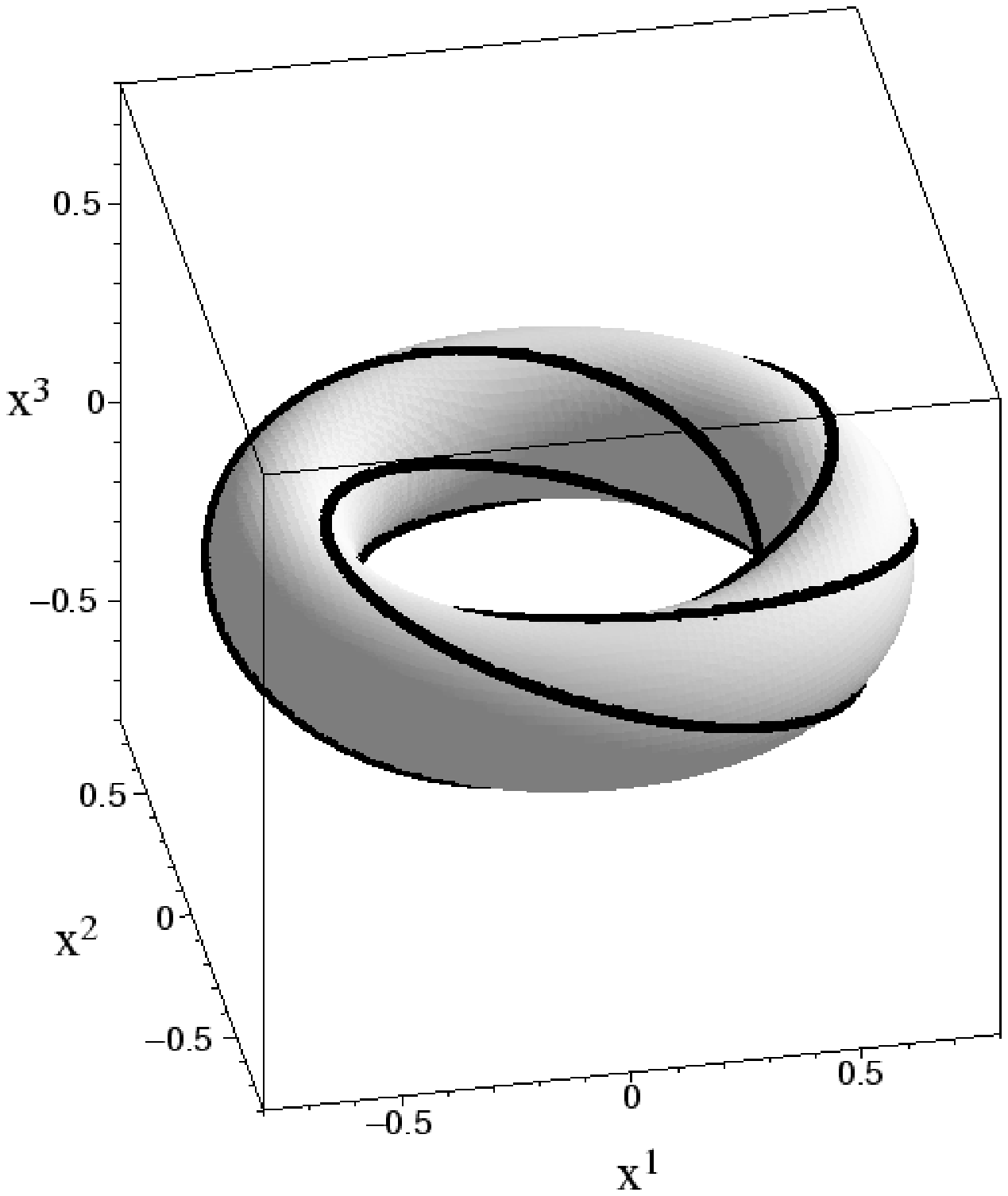}
\end{center}
\end{minipage}
\begin{minipage}[t][][t]{0.45\textwidth}
\hspace*{-2.3cm}\parbox[t]{10\textwidth}{\parbox[t]{1.35\textwidth}{
\caption{ The vector plot shows the projection of the $\bB^3$-field 
onto the $x^1$-$x^3$-plane.
The arrow lengths are not to scale.
The solid lines are contour levels with values 
$-0.5$, $-1$, $-1.8$ of the plotted
density distribution $r\sin\vartheta A^3_\varphi$.
\label{Figure-B3-field-dens-cont}}}}
\end{minipage}
\hspace{0.08\textwidth}
\begin{minipage}[t][][t]{0.45\textwidth}
\hspace*{-2.3cm}\parbox[t]{10\textwidth}{\parbox[t]{1.3\textwidth}{
\caption{The drawn torus is the surface at which 
  $r\sin\vartheta A^3_\varphi=-1.8$.
  All
  field lines of $\bB^3$ lie on toroidal surfaces described by 
  $r\sin\vartheta A^3_\varphi=const$.
  Four field lines lying on this torus are drawn.
\label{Figure-B3-field-lines-on-torus}}}}
\end{minipage}
\end{figure}

\begin{figure}[b]
\begin{minipage}{0.45\textwidth}
\begin{center}
\includegraphics[width=\textwidth]{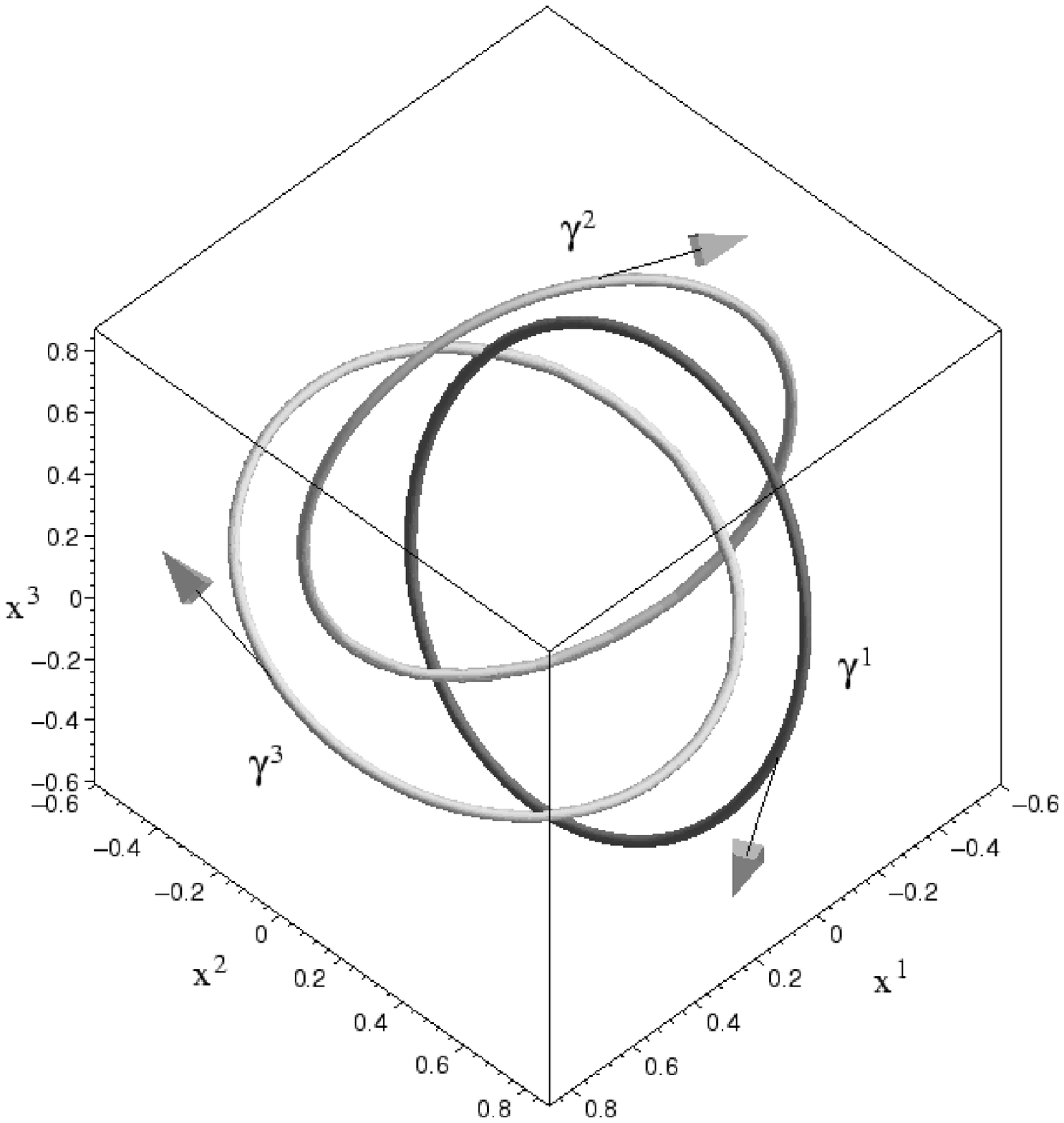}
\end{center}
\end{minipage}
\begin{minipage}{0.45\textwidth}
\begin{center}
\includegraphics[width=\textwidth]{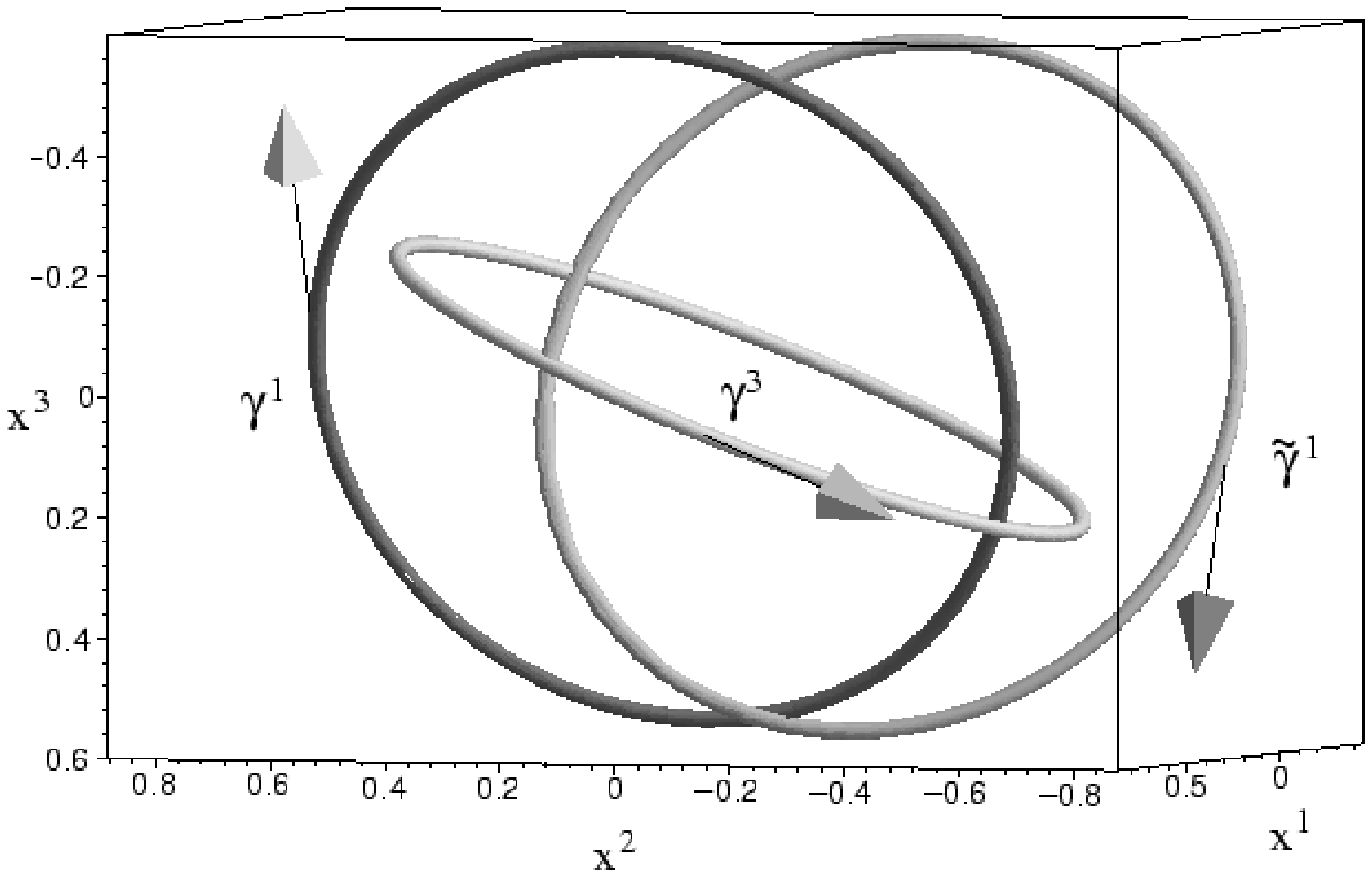}
\end{center}
\end{minipage}
\begin{minipage}{0.45\textwidth}
\hspace*{-2.3cm}\parbox[t]{10\textwidth}{\parbox[t]{1.3\textwidth}{
\caption{The three field lines 
$\gamma^i$ belong to the respective fields $\bB^i$.
Their total linkage is
$\text{lk}(\gamma^1,\gamma^2)+
 \text{lk}(\gamma^2,\gamma^3)+
 \text{lk}(\gamma^3,\gamma^1)=-3$.
\label{Figure-B-1-2-3-field-lines}}}}
\end{minipage}
\hspace{0.08\textwidth}
\begin{minipage}{0.45\textwidth}
\hspace*{-2.3cm}\parbox[t]{10\textwidth}{\parbox[t]{1.3\textwidth}{
\caption{For each field line $\gamma^1$ and $\gamma^3$ 
there exists a field line $\tilde{\gamma}^1$ such that 
  $\text{lk}(\gamma^1,\gamma^3) +
  \text{lk}(\tilde{\gamma}^1,\gamma^3)$ vanishes. 
\label{Figure-zero-cross-helicity}}}}
\end{minipage}
\end{figure}

In our analysis of the three example magnetic fields we now turn our
attention to the topological structure of the fields
 and the linkage of individual field lines.
Figure \ref{Figure-B-1-2-3-field-lines} and
\ref{Figure-zero-cross-helicity}
show numerically integrated field lines, where the starting
points for integration are indicated by the foot points
of the arrows that give the field line direction.
We observe that all field lines are closed and have an elliptical shape.
Figures 
\ref{Figure-B3-field-dens-cont} and 
\ref{Figure-B3-field-lines-on-torus}
visualize the toroidal structure of the
individual fields at the example of $\bB^3$.
Using $\bA^3 =   A^3_r\, \be_r + A^3_\vartheta\, \be_\vartheta + 
                 A^3_\varphi\, \be_\varphi $ 
and $\frac{\p}{\p \varphi}\bA^3=0$ it follows
that $\bB^3$ can be written
\begin{equation}
   \bB^3 = \nabla\times\bA^3 = \frac{1}{r\sin\vartheta}
         \nabla(r\sin\vartheta A^3_\varphi)\times\be_\varphi
                           + B^3_\varphi\be_\varphi. 
\end{equation}
Therefore, the field lines of $\bB^3$ lie on $\varphi$-invariant
toroidal surfaces described by
$r\sin\vartheta A^3_\varphi = \mbox{const}$.
Figure \ref{Figure-B3-field-dens-cont}
shows the poloidal $\bB^3$-field and contour lines for three different
values of $r\sin\vartheta A^3_\varphi$.
A toroidal surface with
$r\sin\vartheta A^3_\varphi = -1.8$ and four field lines on it is
drawn in Figure \ref{Figure-B3-field-lines-on-torus}.
The central field line, sitting within all tori is characterized by
$B^3_r = B^3_\vartheta = 0$. In view of the last equation, this is
equivalent to
$\nabla(r\sin\vartheta A^3_\varphi)=0$ which yields $r=\sqrt{1/3}$
and $\vartheta=\pi/2$.
We observe that all field lines wind around the central field line
exactly once. From this and the toroidal structure of $\bB^3$ we can
conclude that any two arbitrary field lines $l$ and $l'$ 
of $\bB^3$ have a Gau\ss~ linkage $lk(l,l')=1$.

Finally let us discuss the linking properties among field lines of
different fields.
To give an example,
one field line of each field is plotted in
Figure \ref{Figure-B-1-2-3-field-lines}.
The symmetric appearance is
due to the choice of symmetric starting points for the field line
integration.
As was stressed above, the magnetic fields $\bB^i$ can be obtained
from one another by cyclic permutations of the
Cartesian coordinates $x^i$. In the same way the integration starting
points for the field lines $\gamma^1$, $\gamma^2$ and $\gamma^3$ where
chosen to be the cyclic permuted coordinate triples
$(0, 0.8, 0)$,  $(0, 0, 0.8)$ and  $(0.8, 0, 0)$.
It is interesting that the total linkage of the set of field lines is
$\text{lk}(\gamma^1,\gamma^2)+
 \text{lk}(\gamma^2,\gamma^3)+
 \text{lk}(\gamma^3,\gamma^1)=-3$.
Even though we have seen that the mutual
cross-helicities of all three fields vanish, their individual field lines,
in general, are linked pairwise.
To be more precise:
If we e.g.~fix one field line $\gamma^3$ of $\bB^3$, then all
field lines of $\bB^1$ and $\bB^2$ are either linked with
$\gamma^3$ exactly once or they intersect $\gamma^3$ twice.
For reasons of symmetry, there exists for each field line $\gamma^1$
of $\bB^1$ a field line $\tilde{\gamma}^1$ of $\bB^1$,
such that we find the total linkage 
$\text{lk}(\gamma^1,\gamma^3) + \text{lk}(\tilde{\gamma}^1,\gamma^3)=0$.
Hence, the cross-helicity $H(\bB^1,\gamma^3)$ vanishes, which in turn
implies $H(\bB^1, \bB^3)=0$. Figure \ref{Figure-zero-cross-helicity}
shows such field lines
$\gamma^1$, $\tilde{\gamma}^1$ and $\gamma^3$
with $(0, 0.8, 0)$,  $(0, -0.8, 0)$ and  $(0.8, 0, 0)$ as their
respective starting points for the field line integration.
Finally let us remark that in the same way as we obtained
$\tilde{\gamma}^1$ we can obtain field lines
$\tilde{\gamma}^2$ and $\tilde{\gamma}^3$,
here with integration starting points
$(0, -0.8, 0)$,  $(0, 0, -0.8)$ and  $(-0.8, 0, 0)$.
Together, these three field lines
yield a configuration complementary to the one shown in
Figure \ref{Figure-B-1-2-3-field-lines},
which now has a total linkage of $+3$.

\section{Conclusions}

An integral expression has been presented which 
generalizes the third-order invariant
known from the Massey triple product,
to an invariant not limited to mutually unlinked flux tubes,
if the involved fields allow for potentials that obey 
$\nabla\cdot\left(\bA^i\times\bA^j\right)=0$ for~$i\ne j$.
An example shows that the new invariant $H^{(3)}$
is a true generalization.
In our derivation helicity and $H^{(3)}$ emerge from the
same general identity, which involves the Chern-Simons three-form
in the context of an $SU(2)$ gauge symmetry.
Whether this identity leads to further results for other gauge groups 
has not yet been investigated, but it is clear that only expressions
quadratic and cubic in magnetic flux can be obtained.
The constructed invariant is to be seen as 
a ``cross-linkage'' of three fields.
It still remains to clarify whether or how a total third-order invariant 
can be constructed and whether this is possible with 
the help of a cross third-order linkage such as in the case of
helicity.
There might e.g.~exist a subdivision of a single field
into three components such that the total third-order linkage is
determined by the cross-linkage alone.  
Unfortunately, the antisymmetry of
$H^{(3)}(\bB^1,\bB^2,\bB^3)$ seems to
be one of the key problems for a further generalization analogous to
helicity.

\ack The authors gratefully acknowledge financial support from Volkswagen Foundation and  helpful discussions with Mitchell A. Berger.

\appendix
\section{Equivalence of boundary conditions}
\label{Equivboundaryconditions}
We prove for a simply connected volume $V$ the equivalence of the boundary conditions $\bB\cdot\bn|_{\p V}=0$ and $\bA\times\bn|_{\p V} = 0$.

First note that $\bA\times\bn|_{\p V} = 0$ implies $\bB\cdot\bn|_{\p V}=0$: Locally on the boundary we can write $\bn=\nabla \beta$, where $\beta(\bx)$ is a scalar function defined such that $\beta(\bx)= \beta_0$ defines the boundary $\partial V$.
Thus $\bA\times\bn|_{\p V}=0$ implies $\bA|_{\p V} = \alpha \nabla\beta$ for some $\alpha(\bx)$.
Then $\bB|_{\p V}=\nabla\alpha\times\nabla\beta$ and therefore
$\bB\cdot\bn|_{\p V}=0$.

To prove the reverse we start with an arbitrary vector potential $\bA$   
which will in general have a non-vanishing component $\bA_{||}$
tangential to the surface $\p V$.
We can express $\bA_{||}$ as a one-form $\alpha$ 
defined only on $\p V$.
Then the assumption 
$\bB\cdot\bn|_{\p V} = (\nabla\times\bA_{||})\cdot\bn = 0$ 
written in differential forms reads
$d\alpha=0$ on $\p V$.
From $V$ being simply connected it follows that 
$\p V$ has the same homotopy type as the two-sphere $S^2$.
But since the cohomology vector space $H^1(S^2;\reals)=0$,
all closed one-forms are exact.
Therefore there exists a scalar function $\psi$ on $\p V$ such that
$\alpha = d\psi$.
This in turn implies that a gauge exists such that
$\bA_{||}|_{\p V}=0$ and thus $\bA\times\bn|_{\p V}=0$.

%%%%%%%%%%%%%%%%%%%%%%%%%%%%%%%%%%%%%%%%%%%%%%%%%%%%%%%%%%%%%%%%%%
%%%%%     Appendix: Gauge invariance
%%%%%%%%%%%%%%%%%%%%%%%%%%%%%%%%%%%%%%%%%%%%%%%%%%%%%%%%%%%%%%%%%%
\section{Gauge invariance of the third-order invariant $H^{(3)}$}
\label{appendix-H-3-gauge-invariance}
We now prove that the
integral invariant (\ref{H-3-continuous-integral}) is unchanged under 
all gauge transformations
$\bA^i \rightarrow {\bA^i}' = \bA^i + \nabla\phi^i$ for $i=1,2,3$,
which obey the following two conditions:
First we require
\begin{equation}
   \nabla\cdot\bG^i = \nabla\cdot{\bG^i}' = 0, \label{gauge-condition-1}
\end{equation}
where as before we define
$\bG^i = \bA^j \times\bA^k$ and ${\bG^i}' = {\bA^j}' \times{\bA^k}'$
for cyclic indices $i,j,k$.
Second, the gauge transformations must respect the boundary condition
\begin{equation}
   \bA^i \times \bn|_{\p V} = {\bA^i}' \times \bn|_{\p V} = 0, 
                                                \label{gauge-condition-2}
\end{equation}
where $\bn$ is a normal vector to the boundary $\p V$ of the
integration volume $V$.

It is easily checked that a general gauge transformation that leaves the
condition
   $\nabla\cdot\bG^i = 0$
unchanged for $i=1,2,3$
has to be a simultaneous gauge transformation of all three fields. 
Substituting  $\bA^i + \nabla \phi^i$ for $\bA^i$ our invariant of
\Eq (\ref{H-3-continuous-integral}) changes according to
\begin{eqnarray}
\fl
H^{(3)} \rightarrow {H^{(3)}}' 
     =  \int_V (\bA^1 + \nabla \phi^1)\cdot
             [(\bA^2 + \nabla \phi^2)\times(\bA^3 + \nabla   \phi^3)]\dxxx \\
     \eql  \int_V \bA^1 \!\cdot (\bA^2 \times \bA^3) \dxxx \nonumber\\
     +\!\int_V\! \bigl[    \nabla\phi^1\!\cdot(\bA^2 \times \bA^3) 
                     + \nabla\phi^2\!\cdot(\bA^3 \times \bA^1) 
                     + \nabla\phi^3\!\cdot(\bA^1 \times \bA^2) \bigr]\dxxx \nonumber\\ 
     +\!\int_V\! \bigl[   \bA^1\!\cdot( \nabla \phi^2 \times \nabla \phi^3)
                    + \bA^2\!\cdot( \nabla \phi^3 \times \nabla \phi^1)
                    + \bA^3\!\cdot( \nabla \phi^1 \times \nabla\phi^2)
                                                                  \bigr]\dxxx\nonumber\\
     +\!\int_V\!  \nabla \phi^1\!\cdot(\nabla \phi^2 \times \nabla \phi^3) \dxxx.\nonumber
\label{hcs_gauge_transform}
\end{eqnarray}
{\sloppy
We have to show that for gauge transformations respecting 
equations (\ref{gauge-condition-1}) and (\ref{gauge-condition-2}),
\mbox{${H^{(3)}}' =  H^{(3)}$},
i.e. the sum of the last three integrals in
\Eq (\ref{hcs_gauge_transform})  has to vanish.
We can rewrite the first integral as
\begin{eqnarray}
\int_V      \nabla\phi^i\cdot\bG^i \dxxx
\,=\,
        \int_{\p V} \phi^i\bG^i\cdot\bn\dxx     
        \,-\,\int_V      \phi^i(\nabla\cdot\bG^i)
        \dxxx
\,=\,  0,\nonumber
\end{eqnarray}
which vanishes 
since $\bA^i\times\bn|_{\p V}=0$ implies $\bG^i\cdot\bn|_{\p V}=0$.
}% end sloppy
To show that the second integral vanishes we use the identities
\begin{eqnarray}
\nabla \phi^2 \times \nabla \phi^3 &\,=\,\nabla\times(\phi^2 \nabla \phi^3) 
                                    \,=\,-\nabla\times(\phi^3 \nabla \phi^2)\\
&\,=\, \frac{1}{2}\nabla\times( \phi^2 \nabla \phi^3 - \phi^3 \nabla \phi^2) 
\label{first-identity}
\end{eqnarray}
and \[ \nabla\cdot{\bG^1}'-\nabla\cdot\bG^1 = 
            \nabla\phi^3\cdot\bB^2 -\nabla\phi^2\cdot\bB^3, \]
as well as expressions obtained by cyclic 
permutations of the indices $(1,2,3)$.
Substituting the first identity into the second integral we obtain
\begin{eqnarray}
 \fl
 \int_V\Bigl[
   \bA^1\cdot(\frac{1}{2}\nabla\times(\phi^2\nabla\phi^3-\phi^3\nabla\phi^2))
            + (1,2,3)~\text{cyclic}\Bigr]\dxxx \nonumber \\
   \eql-\frac{1}{2}\int_{\p V} \Bigl[
    (\bA^1 \times (\phi^2\nabla\phi^3-\phi^3\nabla\phi^2))\cdot\bn
      + (1,2,3)~\text{cyclic} \Bigr]\dxx\nonumber\\ 
    \vphantom{-}+\frac{1}{2}\int_V \Bigl[ (\nabla \times\bA^1)\cdot
    (\phi^2\nabla\phi^3-\phi^3\nabla\phi^2)+(1,2,3)~\text{cyclic}\Bigr]\dxxx\nonumber\\
   \eql +\frac{1}{2}\int_V \Bigl[\phi^1(\bB^3\nabla\phi^2-\bB^2\nabla\phi^3)
                                           +(1,2,3)~\text{cyclic}\Bigr]\dxxx\nonumber\\
 \eql - \frac{1}{2}\int_V \Bigl[\phi^1(\nabla\cdot{\bG^1}'-\nabla\cdot\bG^1)
                                           +(1,2,3)~\text{cyclic}\Bigr]\dxxx\nonumber\\
 \eql \phantom{-}0.\nonumber
\end{eqnarray}
The surface integral vanishes due to condition (\ref{gauge-condition-2})
and the last volume integral due to condition (\ref{gauge-condition-1}).
Finally we can see that the third integral vanishes by
substituting \Eq (\ref{first-identity}) for the term 
$\nabla\phi^2\times\nabla\phi^3$
\begin{eqnarray}
  \fl
   \int_V \nabla \phi^1\cdot(\nabla\phi^2\times\nabla\phi^3)\dxxx
&= \frac{1}{2} 
   \int_V(\nabla \phi^1\cdot\nabla\times(\phi^2\nabla\phi^3-\phi^3\nabla\phi^2)
   )\dxxx\nonumber\\
&= \frac{1}{2}
   \int_{\p V}
   ( \phi^2 \nabla \phi^3-\phi^3\nabla\phi^2)
   \times
   \nabla\phi^1
   \cdot\bn\dxx\nonumber\\
&= 0.\nonumber
\end{eqnarray}
In the last step we used that due to condition
(\ref{gauge-condition-2}) the gradients $\nabla \phi^1$, $ \nabla \phi^2$ 
and $ \nabla \phi^3$ have to be parallel to $\bn$.
This completes the proof.

\section{Equivalence of the link integrals for disjunct flux tubes} \label{Equivalentintegrals}
The equivalence of the third-order link integrals as given by  
Monastyrsky and Sasorov (1987), Berger (1990)
and Ruzmaikin and Akhmetiev (1994) for three disjunct and mutually
unlinked flux tubes $U_i$ with the integral
(\ref{H-3-continuous-integral}) is shown as follows. Monastyrsky and
Sasorov gave an integral which corresponds to the Massey triple product
and reads in vector notation:
\begin{equation}\fl
 \int_{\partial U_1} ({\bf A}_1 \times {\bf F}_1 - {\bf A}_3 \times
 {\bf F}_3) \dxx,~~
 \begin{array}{r}
    \mbox{with $\nabla \times {\bf F}_1 = {\bf G}_1 :=
    \bA^2\times\bA^3$,}
    \mbox{ for 1,2,3 cyclic.}
 \end{array}\label{Massey} 
\end{equation}
The integration is taken over the surface of tube $U_1$. Cyclic
permutations of indices in (\ref{Massey}) yield equivalent
expressions. Note that the ${\bf G}_i$ in this representation are
evaluated only outside the tubes $U_i$ where $\nabla \cdot {\bf
  G}_i=0$ for {\it any} gauge. To convert this integral into a volume
integral over the whole space, one has to evaluate ${\bf G}_i$ and
${\bf F}_i$  on $U_i$ and therefore encounters the problem of $\nabla
\cdot ({\bf A}_j \times {\bf A}_k) |_{U_i} \neq 0 $ for an arbitrary
gauge. To overcome this problem Berger defined ${\bf G}_i$ within the
flux tubes as
\begin{equation} \fl
{\bf G}_i :=  \left\{\begin{array}{l} {\bf A}_j \times {\bf A}_k - \Phi_{(j)k} {\bf B}_j \quad \mbox{on} \quad U_j , \\ 
 {\bf A}_j \times {\bf A}_k  + \Phi_{(k)j} {\bf B}_k\quad \mbox{on} \quad U_k , \\  
 {\bf A}_j \times {\bf A}_k \quad \mbox{else}, \\ 
                   \end{array} \right. 
\quad \mbox{with}\  i,j,k \ \mbox{cyclic and}\ \nabla \Phi_{(i)j}  = {\bf A}_j|_{U_i} \ .  \label{BergerG} 
\end{equation}
The resulting volume integral 
\begin{eqnarray}
\int_V{{\bf B}_1 \cdot\left({\bf F}_1 - \Phi_{(1)2} {\bf A}_3 \right) \ d^3x } \label{Bergerint}
\end{eqnarray}
is equivalent to (\ref{Massey}), as shown in Berger (1990).
Using the same construction with potentials  $\Phi_{(i)j}$ Ruzmaikin and
Akhmetiev (1994) have rewritten (\ref{Bergerint}) in a more symmetric form.

Now, instead of using the additional potentials $\Phi_{(i)j}$, we can
just as well use the special gauge
\[ {\bf A}_i \rightarrow {\bf {\tilde A}}_i := {\bf A}_i -  \nabla \Phi_{(j)i}   -  \nabla \Phi_{(k)i}\ ,     \]
which implies ${\bf {\tilde A}}_i|_{U_j}=0$ for $i \neq j$. In other
words, the corresponding new $\tilde \Phi_{(i)j}$ are set to zero and
definition (\ref{BergerG}) implies condition (i) of the
Theorem in section~\ref{section-third-order-invariant}.

Furthermore, suppressing tildes, (\ref{Bergerint}) turns into
\begin{eqnarray}  \int_V{ {\bf B}_1 \times {\bf F}_1  \ d^3x}  & =  &  \int_V{ {\bf A}_1 \times {\bf G}_1  \ d^3x } + \int_{\partial V}{\left( {\bf A}_1 \times {\bf F}_1 \right) \cdot {\bf n} \ d^2x }\nonumber   \\
& = &   \int_V{ {\bf A}_1 \times {\bf G}_1  \ d^3x } + \int_{\partial V}{ \left({\bf n} \times {\bf A}_1 \right) \cdot {\bf F}_1 \ d^2x } \nonumber \\
& = &   \int_V{ {\bf A}_1 \times {\bf G}_1  \ d^3x } \ ,
\end{eqnarray}
which finally shows the equivalence of the integrals (\ref{Massey}) 
and (\ref{H-3-continuous-integral}) for the case of three pairwise
unlinked flux tubes.

\section*{References}
\begin{harvard}

\item[] Arnol'd V I 1974 {\it Proc. Summer School in Differential
    Equations} (Erevan) Armenian SSR Acad. Sci. 
    [English translation: 1986 {\it Sel.~Math.~Sov.} {\bf 5} 327--45]

\item[] Arnol'd V I and Khesin B A 1998
  {\it Topological Methods in Hydrodynamics}
  Applied Mathematical Sciences vol 125 
  (New York: Springer-Verlag)

\item[] Berger M A 1990  Third-order link integrals 
  \JPA % {\it Journal of Physics A}
  {\bf 23} 2787--93

\item[] Brown M R, Canfield R C and Pevtsov A A (eds)  1999
  {\it Magnetic Helicity in Space and Laboratory Plasmas}
  Geophysical Monographs vol~111
  (Washington: American Geophysical Union)

\item[] Elsasser W M  1956 {\it Reviews of Modern Physics} {\bf 28} 135 

\item[] Evans N W and  Berger M A 1992
  A hierarchy of linking integrals 
  {\it Topological aspects of fluids and plasmas}
  Nato ASI Series E vol 218 
  ed H K Moffatt \etal 
  (Dordrecht: Kluwer Academic Publisher) pp~237-48

\item[] Fenn R A  1983  
  {\it Techniques of geometric topology}
  London Mathematical Society Lecture Note Series vol~57
  (Cambridge: Cambridge University Press)

\item[] Frankel T 1997
  {\it The Geometry of Physics. An Introduction.}
  (Cambridge: Cambridge University Press)

\item[] Gau\ss~C F 1867 {\it Werke} vol~5 (G\"{o}ttingen: K\"{o}nigliche
    Gesellschaft der Wissenschaften) p~602 

\item[] Itzykson C and  Zuber J-B 1980
  {\it Quantum Field Theory.}
  (New York: McGraw-Hill)

\item[] Kraines D 1966 Massey higher products 
  {\it Trans. Am. Math. Soc.} {\bf 124} 431-49

\item[] Massey  W S  1958 
  Some higher order cohomology operations
  {\it Symp.~Int.~Topologia Algebraica, Mexico} (UNESCO) pp 145--54

\item[] \dash 1969
  Higher order linking numbers
  {\it Conf. on Algebraic Topology, Univ.~Illinois at Chicago Circle,
    June 1968} 
  ed V Gugenheim pp 174--205\\
  Reprinted in: 1998 {\em J.~of Knot Theory and Its Ram.}
  {\bf 7} No.3 393--414.

\item[] Moffatt H K 1969 {\it  Journal of Fluid Mechanics}
  {\bf 35} 117-29

\item[] Monastyrsky M I and Sasorov P V  1987 
  Topological invariants in magnetohydrodynamics
  {\it Sov. Phys. JETP} {\bf 66} (4) 683-688

\item[] Ruzmaikin  A and  Akhmetiev P 1994
  Topological invariants of magnetic fields, and the effect of
  reconnection
  {\it Phys. Plasmas} {\bf 1} 331-336

\item[] Woltjer L 1958 {\it Proc.~Nat.~Acad.~Sci.} {\bf 44}  489 

\end{harvard}

\end{document}